\documentclass[fleqn,useAMS,usenatbib]{mnras}
\pdfoutput=1
\usepackage{amsmath}
\usepackage{amssymb}
\usepackage{txfonts}

\usepackage[T1]{fontenc}
\usepackage{ae,aecompl}

\usepackage{graphicx}	
\usepackage{multirow}
\usepackage{multicol}

\def\apjl{{ApJ}}%
\def\apjs{{ApJS}}%
\def\aap{{A\&A}}%
\newcommand{\swift}{{\it Swift}}
\newcommand{\integral}{\textit{INTEGRAL}}

\newcommand{\xmm}{{\it XMM-Newton}}

\newcommand{\rxte}{{\it RXTE}}

\newcommand{\fluxcgs}{ergs~s$^{-1}$~cm$^{-2}$}
\newcommand{\lumcgs}{ergs~s$^{-1}$}

\newcommand{\ThisSource}{IGR J01217$-$7257}

\title[IGR J01217$-$7257 Outbursts]{Multi-wavelength observations of the Be/X-ray binary IGR J01217$-$7257 (=SXP 2.16) during outburst}

\author[C. M. Boon et al.]{
C. M Boon$^{1}$\thanks{E-mail: c.m.boon@soton.ac.uk},
A. J. Bird$^{1}$,
M. J. Coe$^{1}$,
R. H. D. Corbet$^{2,3,4}$,
P. A. Evans$^{5}$,
\newauthor
J. A. Kennea$^{6}$,
H. A. Krimm$^{7,8,9}$,
S. G. T. Laycock$^{10}$,
\&
A. Udalski$^{11}$
\\
$^{1}$ Department of Physics \& Astronomy, University of Southampton, University Road, Southampton, SO17 1BJ, UK\\
$^{2}$ University of Maryland Baltimore County, Baltimore, MD 21250, USA\\
$^{3}$ CRESST/Mail Code 662, X-ray Astrophysics Laboratory, NASA Goddard Space Flight Center, Greenbelt, MD 20771, USA\\
$^{4}$ Maryland Institute College of Art, 1300 W Mt Royal Ave, Baltimore, MD 21217, USA\\
$^{5}$ University of Leicester, X-ray and Observational Astronomy Research Group, Leicester Institute for Space and Earth Observation, \\ \, \,Department of Physics \& Astronomy, University Road, Leicester, LE1 7RH UK.\\
$^{6}$ Department of Astronomy and Astrophysics, The Pennsylvania State University, University Park, PA 16802, USA\\
$^{7}$ CRESST/Mail Code 661, NASA Goddard Space Flight Center, Greenbelt, MD 20771, USA\\
$^{8}$ Universities Space Research Association, 10211 Wincopin Circle, Suite 500, Columbia, MD 21044, USA \\
$^{9}$ National Science Foundation \\
$^{10}$ Department of Physics, University of Massachusetts Lowell, MA, 01854, USA\\
$^{11}$ Warsaw University Observatory, Aleje Ujazdowskie 4, 00-478 Warsaw, Poland
}

\date{Accepted XXX. Received YYY; in original form ZZZ}

\pubyear{2016}

\begin{document}
\label{firstpage}
\pagerange{\pageref{firstpage}--\pageref{lastpage}}
\maketitle

\begin{abstract}
We present simultaneous, multi-wavelength observations of the Small Magellanic Cloud Be/XRB \ThisSource\,(=SXP 2.16) during outbursts in 2014, 2015 and 2016. We also present the results of \rxte\, observations of the Small Magellanic Cloud during which the source was initially discovered with a periodicity of 2.1652$\pm$0.0001 seconds which we associate with the spin period of the neutron star. A systematic temporal analysis of long term \swift/BAT data reveals a periodic signal of 82.5$\pm$0.7\,days, in contrast with a similar analysis of long base line OGLE {\it I}-band light curves which reveals an 83.67$\pm$0.05\,days also found in this work. Interpreting the longer X-ray periodicity as indicative of binary motion of the neutron star, we find that outbursts detected by \integral\, and \swift\, between 2014 and 2016 are consistent with Type I outbursts seen in Be/XRBs, occurring around periastron. Comparing these outbursts with the OGLE data, we see a clear correlation between outburst occurrence and increasing {\it I}-band flux. A periodic analysis of subdivisions of OGLE data reveals three epochs during which short periodicities of $\sim$1\,day are significantly detected which we suggest are non-radial pulsations (NRPs) of the companion star. These seasons immediately precede those exhibiting clear outburst behaviour, supporting the suggested association between the NRPs, decretion disk growth and the onset of Type I outbursts.  
\end{abstract}





\section{Introduction}
\label{sec:Intro}

Be/X-ray binaries (BeXRBs) are a subclass of high mass X-ray binaries (HXMBs) comprising a significant fraction of Galactic X-ray binaries \citep{2006A&A...455.1165L}. The compact objects associated with BeXRBs are almost always a neutron star, though there are examples of Be-White Dwarf systems and the discovery of a Be star with a black hole companion by \citet{2014Natur.505..378C} suggested there could be a population of as-yet undiscovered Black Hole BeXRBs. In both neutron star and black hole systems, the eccentric orbit of the compact object brings it close to the the circumstellar disk of the primary star star causing either Type I or II X-ray outbursts \citep{1986ApJ...308..669S,2001A&A...377..161O}. Type I outbursts occur periodically around periastron and are caused by the interaction of the compact object with the circumstellar disk material. Conversely, Type II outbursts are aperiodic and caused by the expansion of stellar material to encompass the binary orbit. These outbursts generally last for multiple compact object orbits. Reviews of X-ray and optical behaviour of these systems can be found in \citet{2009IAUS..256..367C} and \citet{2009IAUS..256..361C}. For a comprehensive review of BeXRBs see \citet{2011Ap&SS.332....1R}. For a recent review of HMXBs see \citet{2015A&ARv..23....2W}.

It has been found that the Small Magellanic Cloud (SMC) hosts a population of HMXBs comparable to our Galaxy. All but one of the known SMC HMXBs are BeXRBs, making the SMC a rich target for monitoring these sources, as they provide a population at known distance (62\,kpc, \citealt{2012AJ....144..107H}) and low Galactic obscuration. The SMC was monitored extensively by \rxte\, in the decade from 1999. More recently, the SMC has been regularly monitored by \integral\, as part of a Large Programme. Over the course of an observation period (one year), the SMC was be observed for a total of $\sim$1\,Ms. This programme has had major success in the discovery and classification of new X-ray binaries in the SMC \citep{2010MNRAS.406.2533C,2010MNRAS.403..709M,2015MNRAS.447.2387C}.

In this paper we report on the discovery and follow up observations of the BeXRB, \ThisSource. \ThisSource\, was first discovered by \integral\, during observations of the SMC between 2014 January 11 and 2014 January 12 \citep{2014ATel.5806....1C}. The authors also suggested a potential orbital period of $\sim$84\,days based on visible recurrent outbursts in the {\it I}-band light curve of the 14th magnitude star SMC732.03.3540 from the Optical Gravitational Lenses Experiment (OGLE) Phase IV. Unfortunately, this observation was the last of the SMC during the year-long \integral\, monitoring campaign. Fortunately, the source went into outburst again during the following campaign and it is this outburst that is the main subject of this paper. Using data from OGLE Phase IV, \citet{2014ATel.5889....1S} detected a period of 84.03\,days using a phase dispersion minimisation technique and attributed this to the orbital period of the compact object. 

During \integral\, monitoring of the SMC in 2015, \ThisSource\, was detected in outburst between 2015 October 29 and 2015 October 30 \citep{2015ATel.8246....1C}. Follow-up observations with \xmm\, \citep{Haberl01217} associated the source with the transient X-ray pulsar, SXP2.16 and found the 0.2--10\,keV spectrum well-described by a multi-temperature disk with a power law tail model \citep{2015ATel.8305....1H}. Subsequent analysis of \swift/BAT data by \citet{2015ATel.8312....1C} revealed a peak in the power spectrum at 83.4\,days.

\subsection{\ThisSource\, = SXP 2.16}

SXP 2.16 was first identified during RXTE/PCA observations on 2002 January 5 in which a  transient X-ray pulsar with a periodic signal of 2.1652$\pm$0.0001\,s and flux of 0.625\,mCrab (2--10\,keV) was detected in the direction of the SMC. Pulsations were not detected in a previous observation on 2002 December 13 or follow-up observations on 2003 January 17 \citep{2003IAUC.8064....4C}. \cite{2003ATel..123....1C} suggested the optical counterpart to SXP 2.16 was the emission-line star Lin 526 using spectroscopy from the SAAO 1.9m telescope and searching in the XTE error box, however more recent work has revealed the optical counterpart to be AzV 503, an emission-line star with spectral type B0-5IIe \citep{2014ATel.5806....1C,2015ATel.8305....1H}.

We present analysis of \integral\,, OGLE and \swift\, data taken during the 2014 and 2015 outbursts of \ThisSource\,, archival RXTE observations from the discovery of SXP 2.16 and \swift\, observations of the source during a new outburst in 2016. 
%
\section{Observations}
\label{sec:DataAnalysis}
Details of all observations of \ThisSource\, used in this work are detailed in Table \ref{tab:ObsLog}. The data reduction methods are described below. 
\begin{table*}
\begin{center}
\begin{tabular}{cllcc}
\hline
\hline
Instrument 						& UTC  						 	& MJD 			& Phase 			& Exposure (ks) 		\\
\hline
\rxte\, PCA						& 1999 May 11  15:49:52			& 51309.668		& 0.087				& 4						\\[4pt]
\rxte\, PCA						& 2003 January 05 09:34:00		& 52644.399		& 0.265				& 14					\\[4pt]
\integral ~ (Rev 1373)			& 2014 January 11 00:54:43 --  	& 56668.038 --	& 0.037\,--\,0.051	& 24		 			\\	
 								& 2014 January 12 04:36:28	   	& 56669.192    	&					& 						\\[4pt]
\integral ~ (Rev 1604)			& 2015 October 29 18:34:33 -- 	& 57324.774 --	& 0.997\,--\,0.011	& 21					\\
								& 2015 October 30 20:58:33	 	& 57325.874		&					&						\\[4pt]
\integral ~ (Rev 1606)			& 2015 November 04 01:00:28 --	& 57330.042 -- 	& 0.061\,--\,0.077	& 25					\\
								& 2015 November 05 08:02:24	 	& 57331.335   	& 					& 						\\[4pt]	
\integral ~ (Rev 1608)			& 2015 November 08 11:09:36 --	& 57334.465 -- 	& 0.070\,--\,0.097	& 21					\\
								& 2015 November 09 17:12:28	 	& 57335.717   	& 					& 						\\[4pt]
\integral ~ (Rev 1612)			& 2015 November 20 10:36:28 --	& 57346.442 -- 	& 0.260\,--\,0.270	& 16.5					\\
								& 2015 November 21 06:44:38	 	& 57347.281  	& 					& 						\\[4pt]
OGLE							& 2001 June 30 10:18:32 --		& 52090.430 -- 	& $-$				& $-$ 					\\
								& 2016 September 16 05:18:14	& 57647.221		& 					&						\\[4pt]	
\swift/BAT ~ (Back-processed)	& 2005 February 06 --			& 53407 -- 		& $-$				& $-$					\\
								& 2014 February 20	 			& 56708   		& 					& 						\\[4pt]	
\swift/BAT  (Transient Monitor)	& 2014 January 22 --			& 56679 -- 		& $-$				& $-$					\\
								& 2016 October 13	 			& 57674   		& 					& 						\\[4pt]	
\swift/XRT ~ (S-CUBED)			& 2016 June	08 08:19:41	--		& 57547.347 --	& $-$				& $-$					\\
								& 2016 August 31 11:02:24		& 57631.460		&					&						\\[4pt]
\hline
\hline
\end{tabular}
\caption{Log of observations discussed in this work. OGLE data used here covers both the OGLE--III and OGLE--IV observation campaigns. Phases are calculated using the ephemeris and orbital period of MJD 56995 and 82.5\,days derived in this work using \swift/BAT observations. The duration of the S-CUBED survey data is listed here, each exposure during this period is 60 seconds. S-CUBED data of \ThisSource\, is shown in Fig. \ref{fig:S3_lc} and discussed in Section \ref{sec:XRT_results}. Phases and exposures are not listed for either \swift/BAT, \emph{OGLE} or S-CUBED observations as they cover all binary phases. The \rxte\, observation on MJD 51309.668 is a pre-discovery detection of SXP 2.16 found through analysis of archival SMC observations as discussed in Section \ref{sec:RXTE_data}.}
\label{tab:ObsLog}
\end{center}
\end{table*}
%
\subsection{\integral\, data analysis}
\label{sec:INTEGRAL_Data}
\ThisSource\, was detected by \integral/IBIS \citep{2003A&A...411L.131U, 2003A&A...411L...1W} during observations of the SMC taken between MJD 56668.038 and 56669.192 during Revolution 1373 \citep{2014ATel.5806....1C}. We refer to this observation as the 2014 Outburst for the remainder of this work. The source was subsequently detected in further monitoring of the SMC nearly 2 years later between MJD 57324.774 and 57347.281 during Revolutions 1604--1612. We refer to this as the 2015 Outburst henceforth. We processed data from these two detected outbursts using the Offline Science Analysis (OSA, \citealt{2003A&A...411L..53C}) software version 10.2 . Images were created at the Science Window level (Scw; a single pointing of \integral\, lasting approximately 2\,ks) as well as mosaic images on a revolution-by-revolution basis. The journal of all of these observations are presented in Table \ref{tab:ObsLog}. Light curves for each outburst were produced at science window resolution.

\subsection{\swift\, data}
\label{sec:swift_data}

The Burst Alert Telescope on board the \swift\ satellite is described by \cite{2005SSRv..120..143B}.The observations of \ThisSource\ were taken from the \swift/BAT transient monitor \citep{2013ApJS..209...14K} and cover the energy range 15 -- 50 keV. The time range covered by these observations is from MJD 56679 to 57674. The transient monitor data are provided with time resolutions of one orbit and daily averages; we used the orbit light curve for period searches, and the one day averages for plotting the light curve.

Data were selected and analysed in a similar way to that described in \cite{2013ApJ...778...45C}: we used only data for which the quality flag (``DATA\_FLAG'') was set to 0, indicating good quality, and also excluded points with low count rates and extremely small nominal uncertainties. 

The transient monitor data were extended by analysing data collected from the BAT from the start of the mission until \ThisSource\, was added to the monitor catalog. The back-processed light curve differs from the transient monitor in that only daily averages were produced. In addition, the errors on the fluxes are only statistical and do not include systematic errors that are included in the transient monitor data. The back-processed light curve covers 2005 February 6 to 2014 February 20 (MJD 53,407 to 56,708) and so overlaps the transient monitor light curve by approximately 34 days.

The Swift SMC Survey (S-CUBED) is a wide area/short exposure survey of the SMC in X-rays performed by the Swift X-ray Telescope (XRT) \citep{2016ATel.9299....1K}. The survey consists of 142 x 60 second exposure tiled pointings covering the SMC performed approximately weekly. Observations for S-CUBED began on June 8th, 2016 and \ThisSource\, was detected on 2016 July 6, 10 \& 15. The journal of these observations can be found in Table \ref{tab:ObsLog}.

\subsection{RXTE data}
\label{sec:RXTE_data}
The region of the SMC containing SXP 2.16 was observed four times (1999 May 11, 2002 December 13, 2003 January 05, 2003 January 17) by RXTE as part of the long-term SMC X-ray monitoring programme \citep{2008ApJS..177..189G}. Since the RXTE PCA telescope \citep{1996SPIE.2808...59J,2006ApJS..163..401J} had a large non-imaging field of view (1 degree FWHM) there was always the possibility of other sources contributing to the resulting signal. However, the identification of a pulsed signal is unique to a specific source and a clear indication of identification - even if the pulse fraction is difficult to determine. Hence, because \ThisSource\, lies close to the bright persistent source, SMC X$-$1, after the initial discovery observations in 2003 no further pointings were attempted. 

\subsection{OGLE data}
\label{sec:OGLE_data}
Data from the Optical Gravitational Lenses Experiment (OGLE) Phases III and IV \citep{2008AcA....58...69U, 2015AcA....65....1U} were downloaded from the XROM website\footnote{http://ogle.astrouw.edu.pl/ogle4/xrom/xrom.html} and used to compare the long-term {\it I}-band behaviour of the companion star with the long-term hard X-ray activity as monitored by \swift/BAT. Time stamps and magnitudes of the OGLE observations were converted to Modified Julian Date (MJD)  and Jansky (Jy) respectively. The light curves from the combined OGLE--III and OGLE--IV data sets spanning nearly 15 years are shown in Section \ref{sec:OGLE_results}. 

In order to search for the presence of sub-seasonal variations, the combined OGLE light curves were detrended using polynomial fits. The entire OGLE IV data set (MJD 55347.427 --57647.221) cannot be adequately described by a single polynomial fit. We divided the data in to two sections, the first from MJD 55347.427 to 57434.033 and the second from MJD 57560.425 to 57647.221. The first section is well described by a quartic polynomial, while the second is well described by a linear fit. We use these functional forms as adequate descriptions of the overall behaviour of the companion star and decretion disk during these epochs. The OGLE III (MJD 52090.42954 -- 54873.06695) data also cannot be well described by a single polynomial fit. As a result, we fit each season of OGLE III data with separate polynomials to remove seasonal trends. 
\section{Results}
\label{sec:Results}

\subsection{\integral\, results}
\label{sec:IntResults}
\begin{figure}
\begin{center}
\includegraphics[width=0.5\textwidth,natwidth=720,natheight=621]{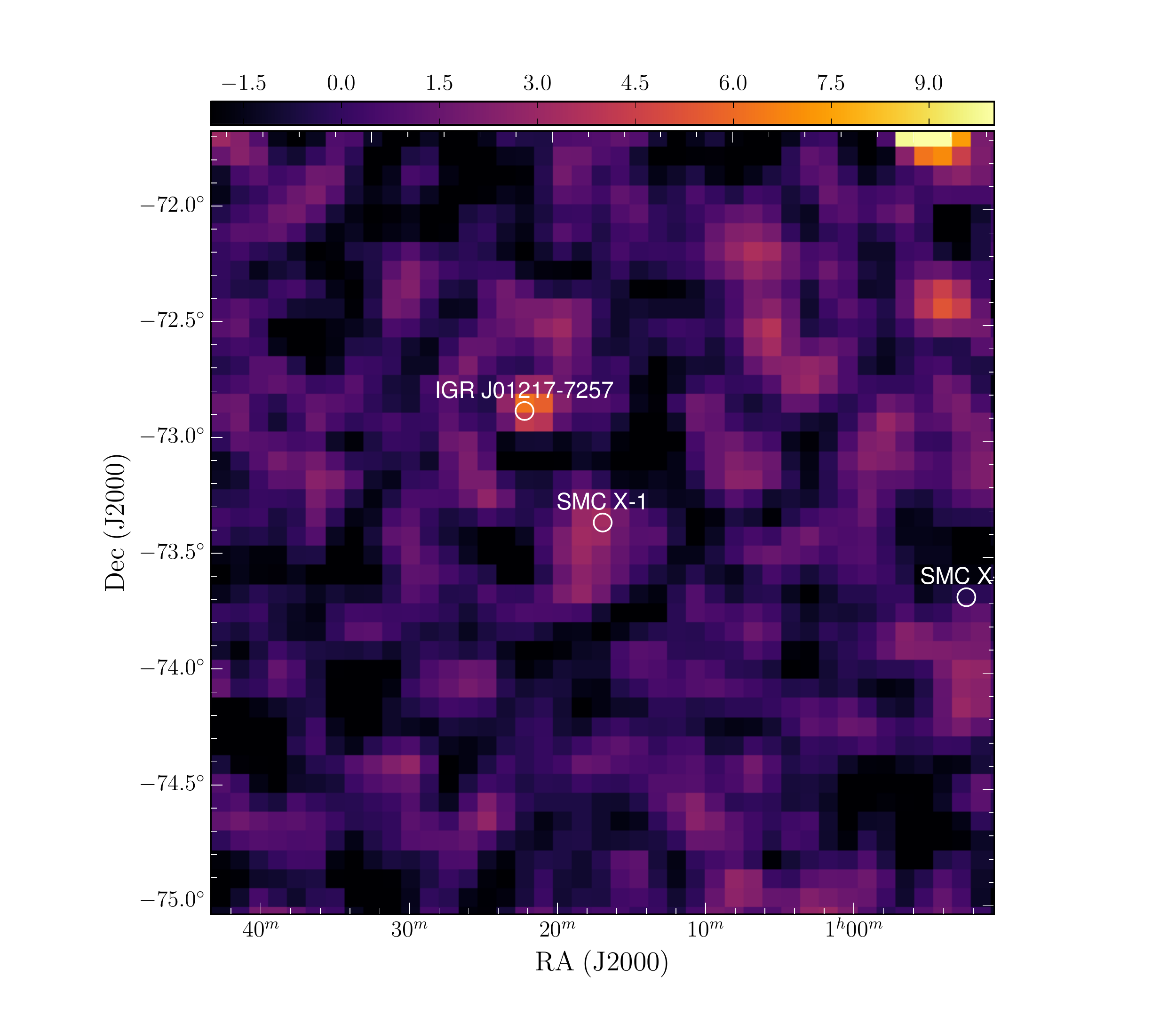}
\caption{IBIS/ISGRI 20--40\,keV significance map of \ThisSource\, observed between MJD 56668.038 and 56669.192. The source is detected at $6.3\sigma$ with a 24\,ks exposure. The best X-ray positions of \ThisSource\,, SMC X$-$1 and SMC X$-$2 are shown in white circles. }
\label{fig:IntMap_2014}
\end{center}
\end{figure}

The results of the \integral\, analysis detailed above are summarised in Table \ref{tab:IBIS_results} and the 20--40\,keV significance map for the 2014 is shown in Fig. \ref{fig:IntMap_2014}. Orbital phases of these observations are calculated using the orbital period and ephemeris derived from X-ray data in Section \ref{sec:BAT_monitorresults}. Conversion of the source intensity from IBIS/ISGRI counts/s to flux was obtained by processing an observation of the Crab taken during Revolution 1597\footnote{Observations used in this work were taken between MJD 57305.334 and 57305.894} in the 20--40\,keV band and using the conversion factors of \citet{2016ApJS..223...15B}. Luminosity calculations were performed assuming a distance to the SMC of 62\,kpc from \citet{2012AJ....144..107H}.

In order to quantify the variability within the science window light curves, we adopt the Bayesian Blocks method of \citet{2013ApJ...764..167S} using a Python implementation\footnote{https://zenodo.org/record/57750\#.V-qBXyMrK37} \citep{adam_hill_2016_57750}. We set the false alarm probability to be 0.003, which corresponds to approximately a 3$\sigma$ change being required before the algorithm accepts there is significant variation between points in the time series.

This analysis has shows that in general, \ThisSource\, shows very little sub-10\,ks  variability and for the most part, shows a statistically constant flux during an \integral\, revolution. However, the source does in general exhibit a decrease in average luminosity between revolutions. We note that the luminosities and orbital phases of Revolution 1373 and 1612 are approximately equal, suggesting that Revolution 1373 is infact the tail end of an outburst. The peak of this outburst was not observed by \integral\, or reported by any other missions.  

\begin{table*}
\begin{center}
\caption{Results from the IBIS/ISGRI analysis of the 2014 and 2015 Outbursts of \ThisSource. The 20--40\,keV flux is given in units of mCrab and (10$^{-11}$\fluxcgs). Conversion between counts and mCrab was obtained using an observation of the Crab taken during Revolution 1597  between MJD 57305.334 and 57305.894 and the conversion factors of \citet{2016ApJS..223...15B}. Luminosities are calculated using a distance of 62\,kpc \citep{2012AJ....144..107H}}
\begin{tabular}{ccccccc}
\hline
\hline
Rev. 	& Intensity 		& Significance 	& Exposure 	& Flux 				& Flux					& Luminosity			\\
		& (cts/s)			& 				& (ks)		& (mCrab) 			& (10$^{-11}$\fluxcgs) 	& (10$^{37}$\lumcgs)	\\
\hline
1373	& 0.59$\pm$0.09		& 6.3			& 24		& 4.21$\pm$0.67		& 3.19$\pm$0.51			& 1.43$\pm$0.23			\\
1604	& 1.56$\pm$0.10		& 15.2			& 21		& 11.79$\pm$0.76	& 8.92$\pm$0.58			& 4.10$\pm$0.27			\\
1606	& 0.77$\pm$0.08		& 9.1			& 25		& 5.83$\pm$0.60		& 4.41$\pm$0.45			& 2.03$\pm$0.21			\\
1608	& 1.13$\pm$0.09		& 12.4			& 21		& 8.50$\pm$0.67		& 6.43$\pm$0.51			& 2.96$\pm$0.24			\\
1612	& 0.57$\pm$0.10		& 5.7			& 16.5		& 4.29$\pm$0.76		& 3.25$\pm$0.58			& 1.50$\pm$0.28			\\	 
\hline
\hline
\end{tabular}
\label{tab:IBIS_results}
\end{center}
\end{table*}

\subsection{\swift\, results}
\label{sec:Swift_results}

\subsubsection{\swift/BAT Transient Monitor Results}
\label{sec:BAT_monitorresults}
To search for and quantify periodic modulation of the hard X-ray data, we calculated the power spectrum of the BAT light curve. This was obtained by calculating the Discrete Fourier Transform (DFT) of the orbital light curve with the contribution of each data point weighted using the ``semi-weighting'' method which accounts for both the error on each data point and the excess variance of the entire light curve \cite{2007PThPS.169..200C,2007ApJ...655..458C}.

\begin{figure}
\begin{center}
\includegraphics[width=0.5\textwidth,natwidth=657,natheight=535]{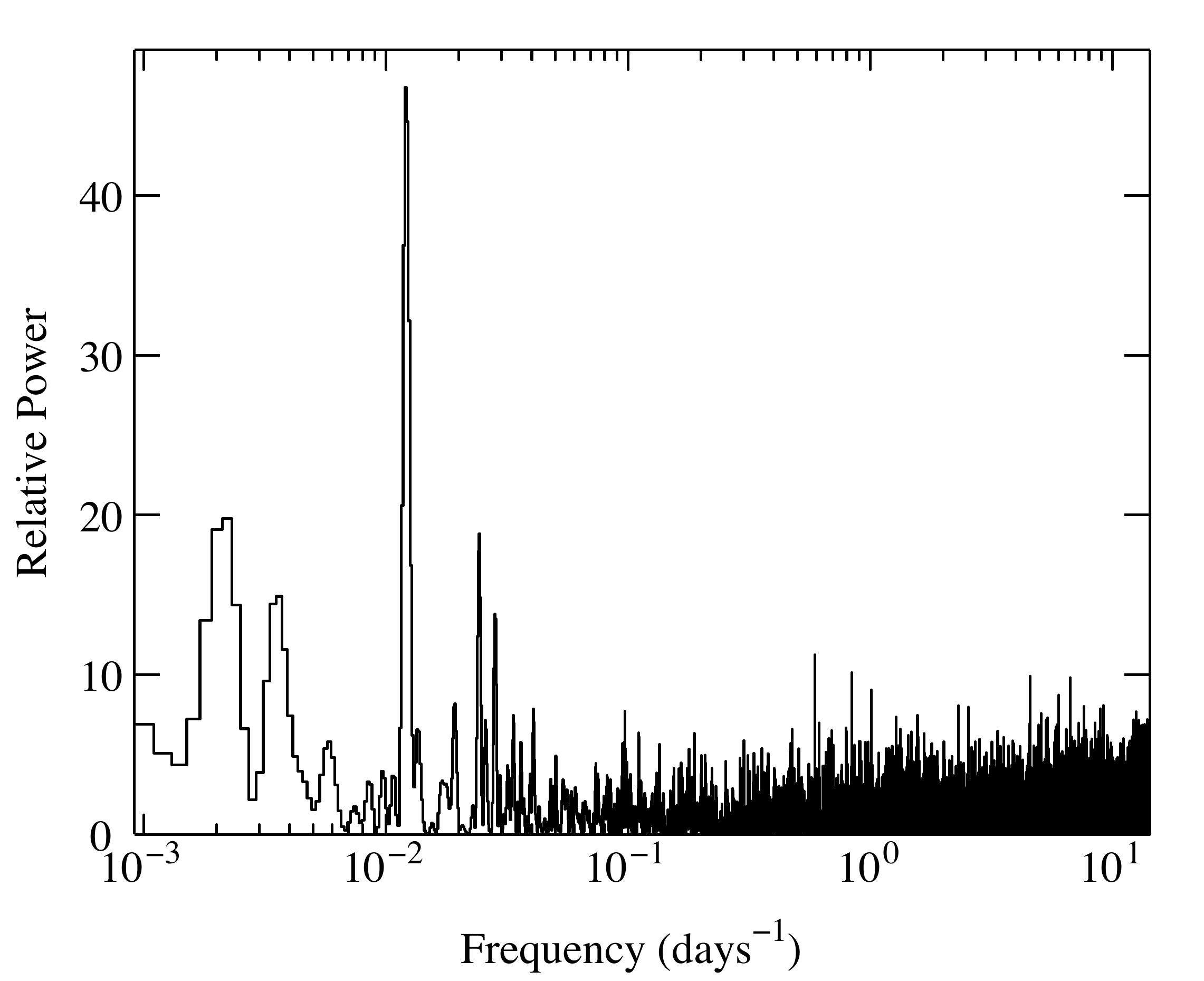}
\caption{Power spectrum of the \swift/BAT light curve of \ThisSource. The strongest peak is at the proposed 82.5 day orbital period.}
\label{fig:bat_power}
\end{center}
\end{figure}

The period range searched was from 0.07 days to the length of the data set (994 days) oversampling the power spectrum by a factor of 5 compared to the nominal Fourier resolution of 1/994 days$^{-1}$. The resulting power spectrum is shown in Fig. \ref{fig:bat_power}. The largest peak in the power spectrum is at a period of 82.5 $\pm$ 0.7 days, where the uncertainty is derived from the prescription of \citet{1986ApJ...302..757H}. There is also a smaller peak at a frequency consistent with the second harmonic of the 82.5 day modulation.

\begin{figure}
\begin{center}
\includegraphics[width=0.5\textwidth,natwidth=734,natheight=536]{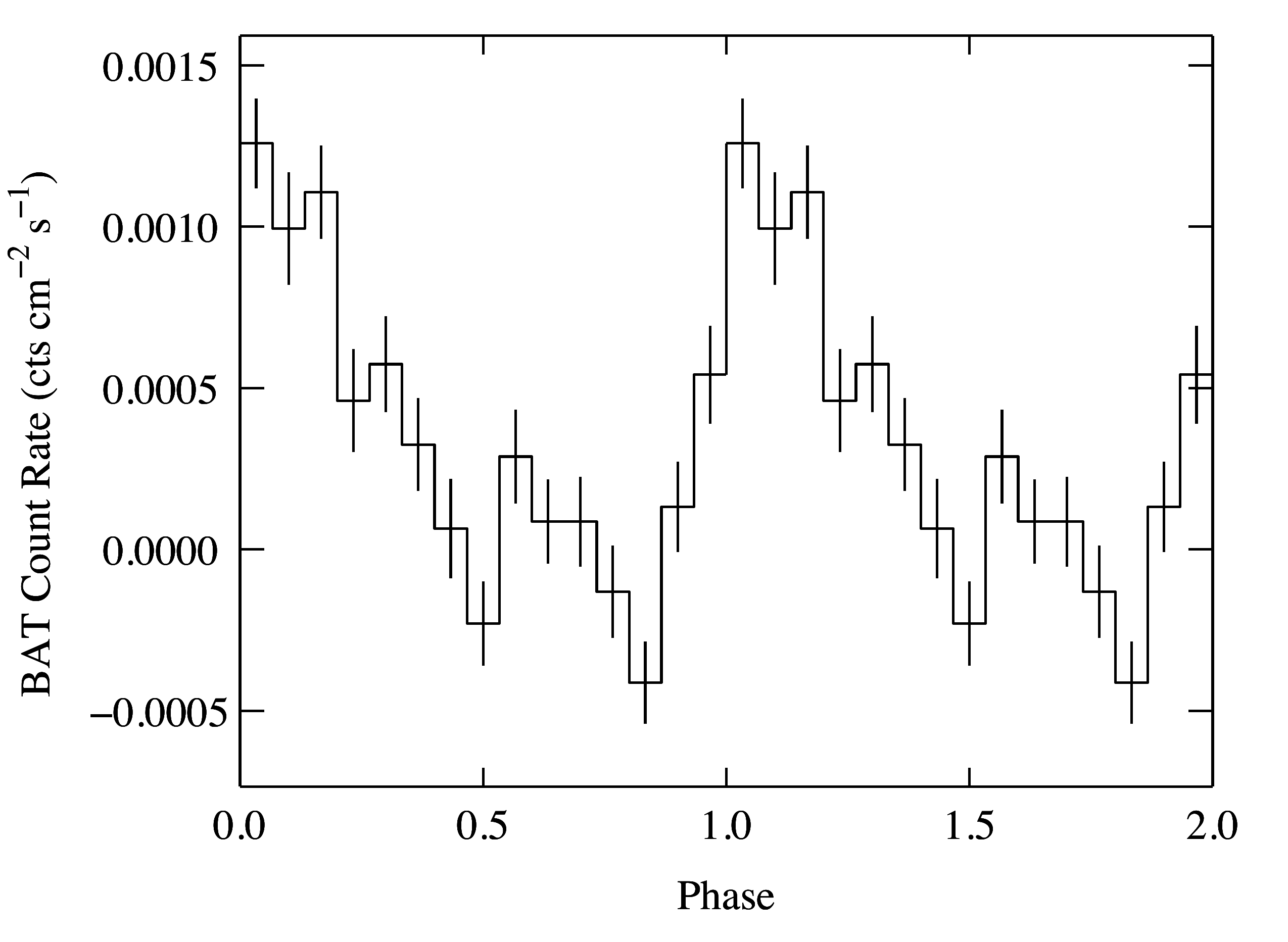}
\caption{\swift/BAT light curve of \ThisSource\, folded on the proposed 82.5 day orbital period with phase zero at MJD 56995.}
\label{fig:bat_fold}
\end{center}
\end{figure}

The BAT light curve folded on the derived period is shown in Fig. \ref{fig:bat_fold}. 

\subsubsection{Back-processed BAT Observations}
\label{sec:BATbackprocess}

We calculated the DFT of the back-processed light curve alone, covering a period range from 2 days to the length of the data set (3302 days) This did not show the presence of any modulation at the 82.5 day period. A direct inspection of the light curve also did not directly reveal large outbursts at the expected times.

We next investigated the combined transient monitor and back-processed light curve.
For consistency we used only the one day resolution transient monitor data and
considered only statistical errors. We excluded the back-processed data that overlapped with the transient monitor data. We then calculated power spectra for periods longer than 2 days and evaluated the height of modulation near 82.5 days compared to the mean power level. The light curves used for this always had the same end time of  2016 October 13 (MJD 57674) but differing start times.

\begin{figure}
\includegraphics[width=0.5\textwidth,natwidth=677,natheight=530]{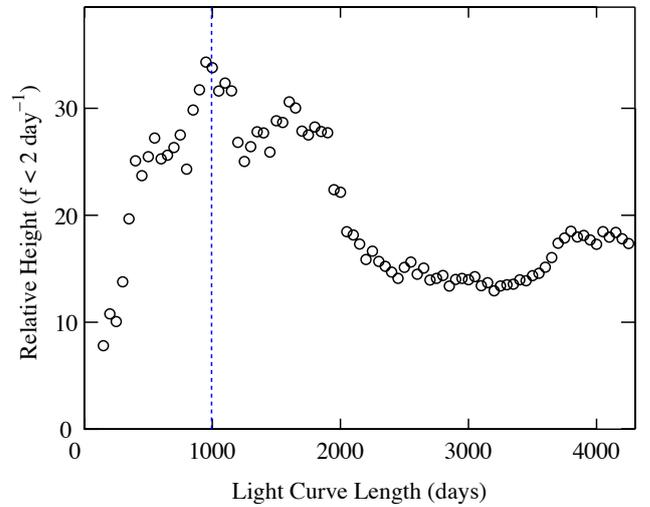}
\caption{Relative height of periodic modulation in power spectra of \swift/BAT light curves of \ThisSource. The light curves used have end dates of MJD 57674. The start date is MJD 57674 minus the light curve length. The vertical dashed blue line indicates the start of the transient monitor data.
}
\label{fig:bat_rvt}
\end{figure}

From Fig. \ref{fig:bat_rvt} it can be seen that the relative height of the modulation initially increased as the start time of the light curve was moved earlier. A maximum in the relative height is found that coincides with the start of the transient monitor data. The relative height then generally declines, with a more significant drop for times starting earlier than 1950 days before the end date. (i.e. for light curves with start times before $\sim$ MJD 55274 =
2010 March 19). As the start time is moved earlier there is a modest increase for a start time of $\sim$ 3800 days before the end date (i.e. for a light curve with start time of $\sim$ MJD 53874 = 2006 May 19) suggesting possible activity at that time.

\begin{figure}
\includegraphics[width=0.5\textwidth,natwidth=682,natheight=544]{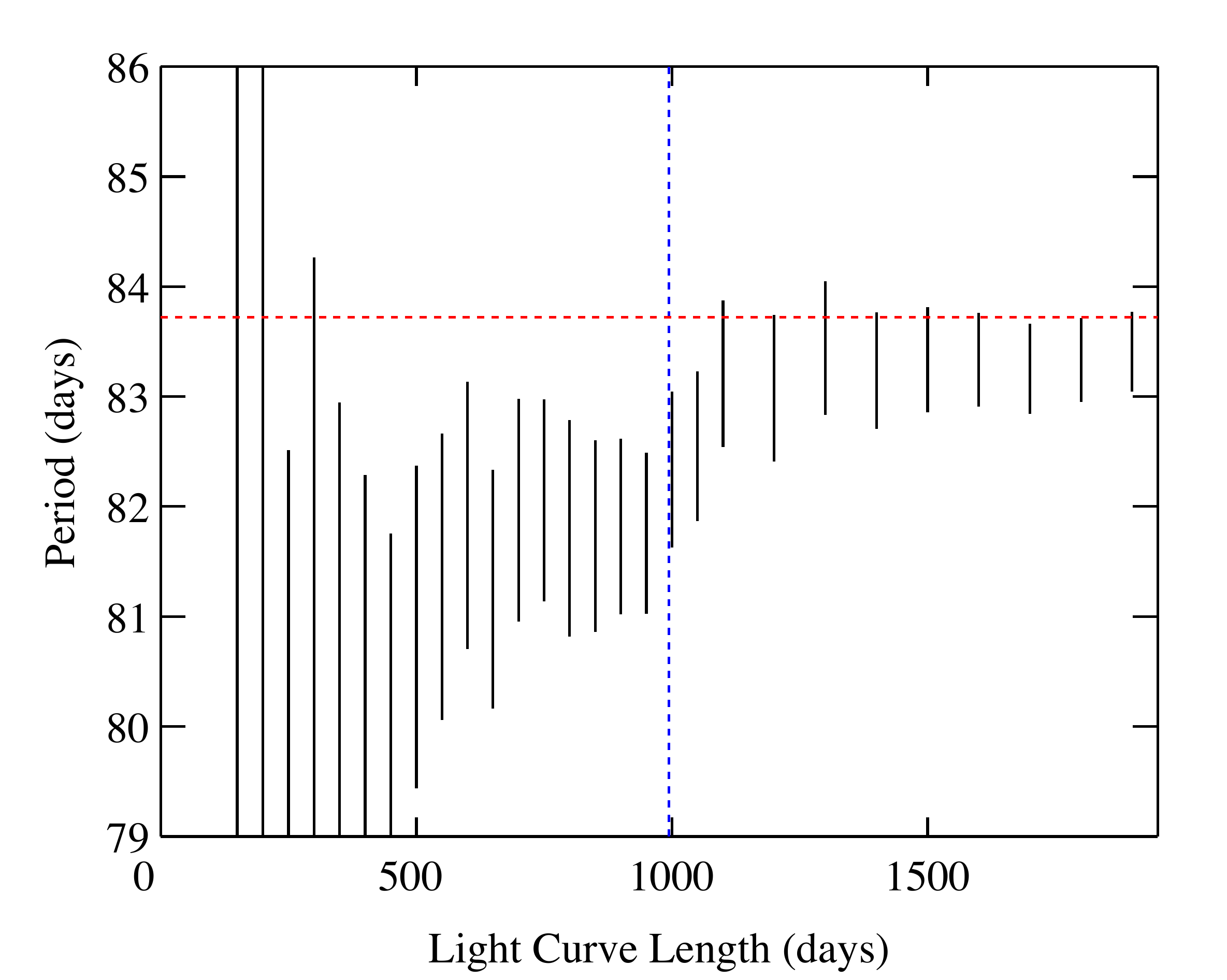}
\caption{Period of the strongest peak in power spectra of
\swift/BAT light curves of \ThisSource. The light curves used had end dates of MJD 57674. The start date is that date minus the light curve length.
The vertical dashed blue line indicates the start of the transient monitor data. The horizontal dashed line indicates the period determined from OGLE observations in Section \ref{sec:OGLE_results}.
}
\label{fig:bat_pvt}
\end{figure}

In Fig. \ref{fig:bat_pvt} we also plot the apparent period determined for light curves of different lengths, all with the same end time as above. We note that the inclusion of data from before the transient monitor data results in an apparently somewhat longer period of approximately 83\,days, that is closer to that determined from the OGLE light curve.

\subsubsection{\swift/XRT Results}
\label{sec:XRT_results}

\begin{figure}
\begin{center}
\includegraphics[width=0.5\textwidth,natwidth=576,natheight=432]{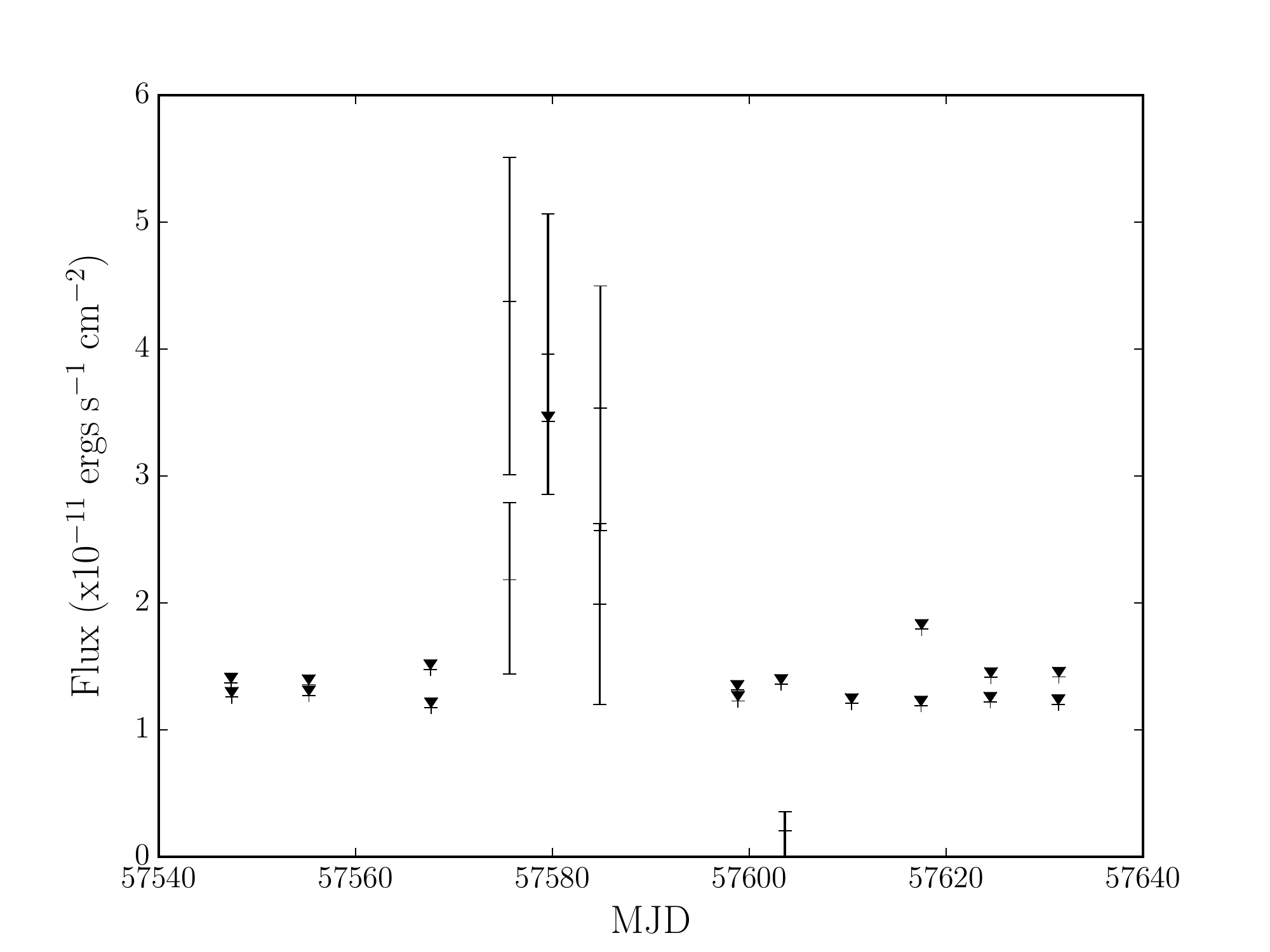}
\caption{0.3--10\,keV light curve of \ThisSource\, as observed by \swift/XRT as part of the S-CUBED survey. There are clear detections on MJD 57575, 57579 and 57584. The peak flux during this outburst is (4.4$\pm$1.2)$\times10^{-11}$\fluxcgs\, corresponding to a luminosity of (2.0$\pm$0.5)$\times10^{37}$\lumcgs\, at 62\,kpc.}
\label{fig:S3_lc}
\end{center}
\end{figure}

One of the sources detected in outburst during the S-CUBED monitoring campaign was SXP 2.16. Fig. \ref{fig:S3_lc} shows the light curve history for this source with clear detections on MJD 57575, 57579 and 57584. This outburst peaked at a flux of (4.4$\pm$1.2)$\times10^{-11}$\fluxcgs\, in the energy range 0.3--10 keV, which corresponds to L$_{\mathrm{x}}$ = (2.0$\pm$0.5)$\times10^{37}$\lumcgs\, for a source in the SMC at a distance of 62\,kpc \citep{2012AJ....144..107H}. The XRT count rate was too low to justify triggering a longer WT observation so we have no measurement of the pulse period during this outburst.
The duration of the outburst ($\sim$9\,d) and the lack of a detection on either side, strongly suggests this is a typical Type I outburst from this system. Calculating the phase of the outburst based upon the period and ephemeris determined in Section \ref{sec:BAT_monitorresults} gives a phase range of $\phi$\,=\,0.030--0.139, confirming the Type I nature of this outburst. 

The association between SXP 2.16 and \ThisSource\, was not immediately clear. Using the spin period as a unique identifier, the same signal was detected in an \xmm\, observation by \cite{2015ATel.8305....1H} at RA\,$=$\,01:21:40.5, Dec\,$=$\, -72:57:32 with an error circle of 5 arcsec. This signal position is consistent the best determined X-ray position of RA\,$=$\,01:21:40.6, Dec\,$=$\,-72:57:21.9 (error circle = 1.5 arcmin) from the \integral/JEM-X detection between MJD 56668.038 and 56669.192, which confirmed the association of these two sources.

\subsection{RXTE results}
\label{sec:RXTE_results}
The RXTE observations of 5 January 2003 revealed the presence of a new periodic signal at 2.165\,s. This was identified as a previously unknown X-ray pulsar and given the designation XTE J0119-731 \citep{2003IAUC.8064....4C}. Other observations, 3 weeks before (13 December 2002) and 2 weeks afterwards (17 January 2003), failed to detect the pulsations. However, we note that the strong signal from the nearby source SMC X-1 significantly reduced the sensitivity, especially since the pulse periods are relatively close together (SMC X-1 is 0.7s). 

\begin{figure}
\begin{center}
\includegraphics[width=0.5\textwidth,natwidth=576,natheight=432]{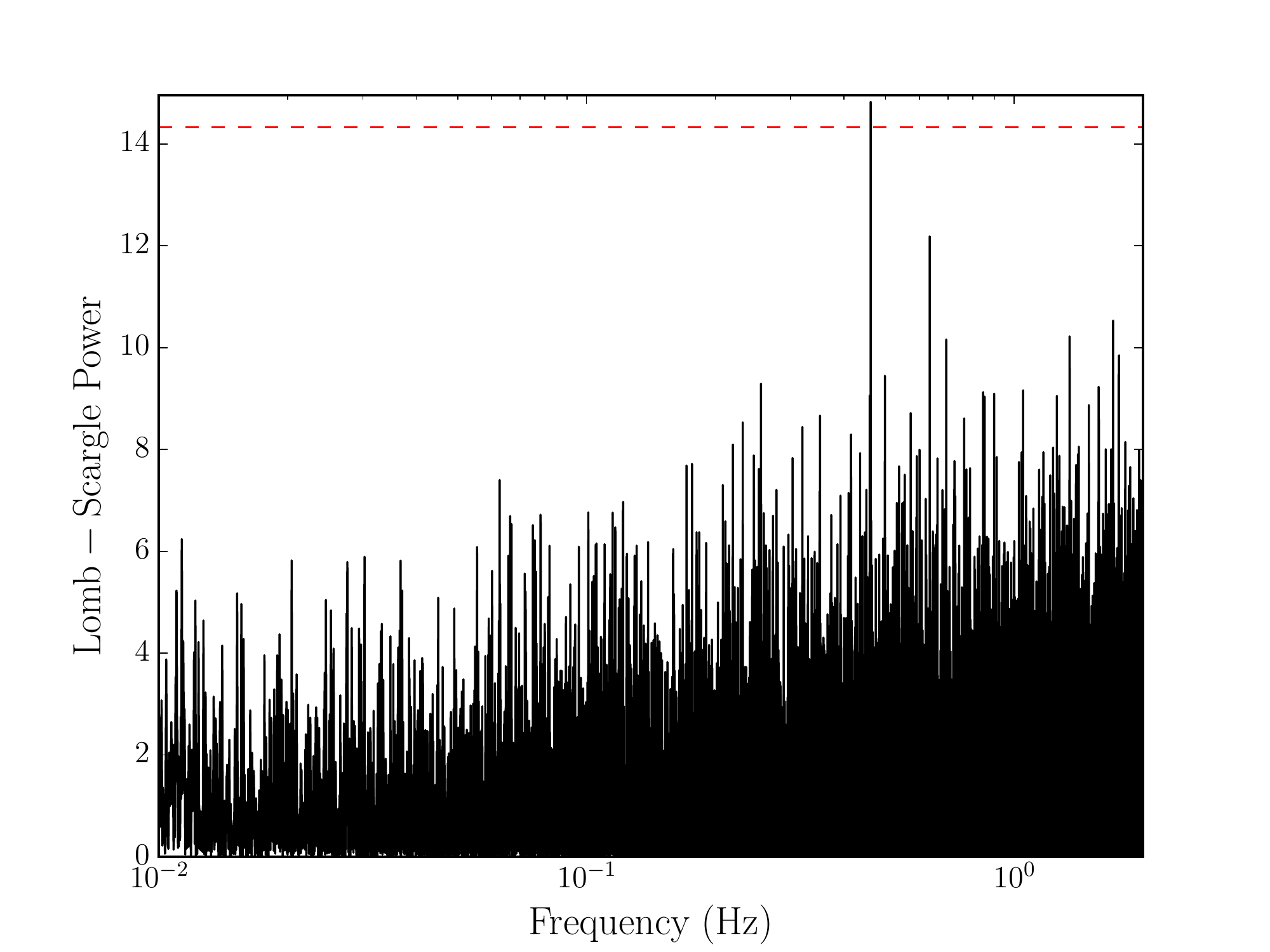}
\caption{Lomb-Scargle Periodogram of data taken by RXTE during an observation of the SMC on MJD 52644.399. The peak in the power spectrum is at 0.4618\,Hz, corresponding to a period of 2.165\,s. The red dashed line indicates the 95\% confidence level of detection. }
\label{fig:XTE_LS}
\end{center}
\end{figure}

\begin{figure}
\begin{center}
\includegraphics[width=0.5\textwidth,natwidth=605,natheight=513]{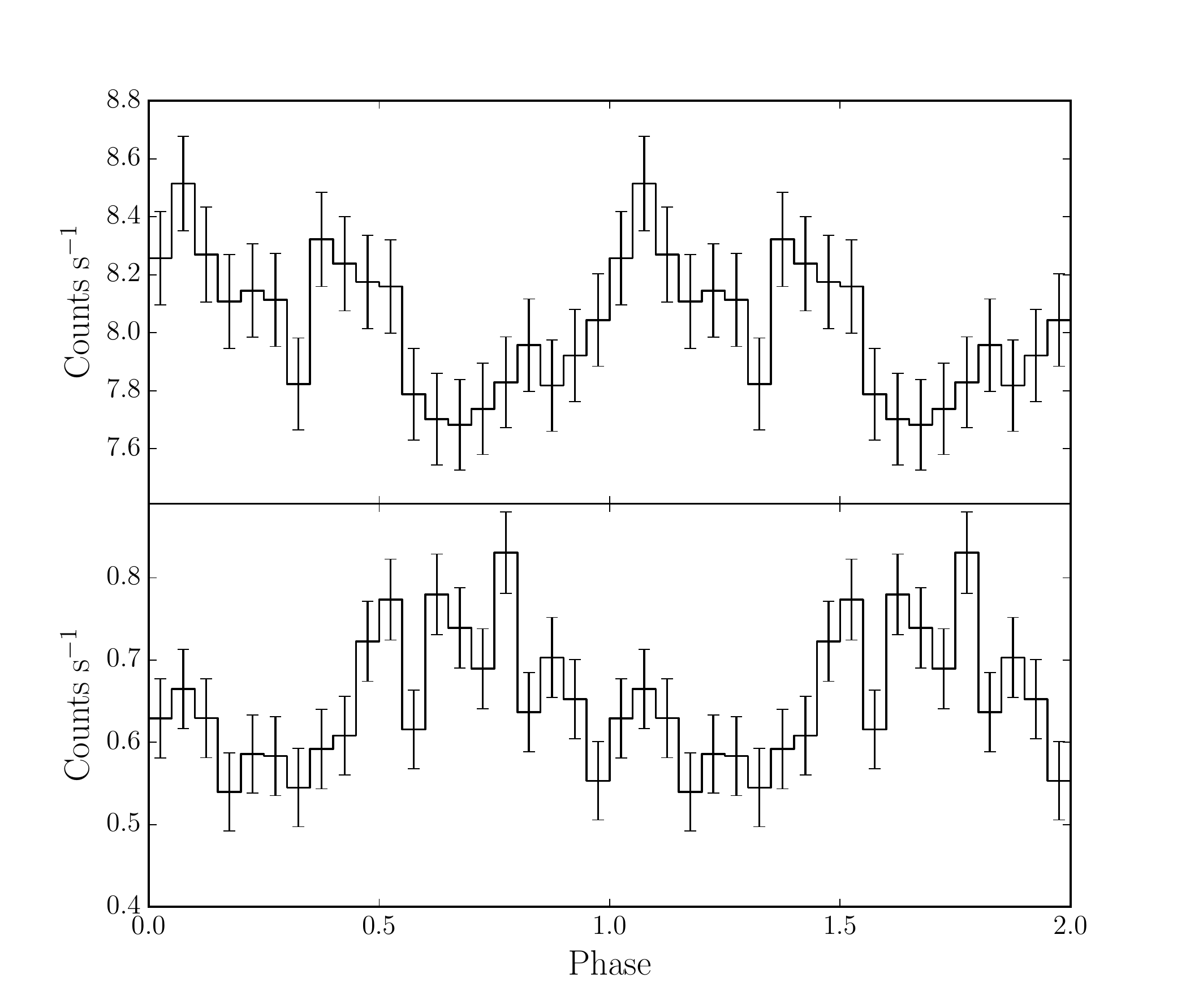}
\caption{Top Panel: Pulse profile of SXP 2.16 from the observation of the SMC taken on MJD 51309.668 by \rxte. The data are folded on a period of 2.164971\,s. Bottom Panel: Pulse profile of SXP 2.16 from the observation of the SMC taken on MJD 52644.399 by \rxte. The data are folded on a period of 2.165\,s and show a complex pulse profile.}
\label{fig:XTE_phasefold}
\end{center}
\end{figure}

Fig. \ref{fig:XTE_LS} shows the power spectrum obtained from the RXTE observation of 2003 January 05. There is a clear strong peak at the fundamental period of 2.165s, with significant power also present in the harmonics. This immediately suggests a complex pulse profile and that is confirmed by folding the data at the fundamental period - see Fig. \ref{fig:XTE_phasefold}, bottom panel for this profile. This complexity in the pulse profile was commented upon by \citet{2015ATel.8305....1H} from their XMM observations during the recent outburst. It is possible that the profile is more complex during the 2015 outburst than the 2003 one because those authors report more power in the harmonics of the power spectrum than in the fundamental. However, it is also possible that the complexity seen in the XMM observation is due to the increased S/N in the data. These complex pulse profiles presumably arise from several emission regions contributing to the X-ray signal.

Subsequent re-analysis of archival \rxte\, data using the PUMA/ORCA pipelines revealed a periodic signal of 2.16497$\pm$0.00014\,s (see \citealt{2008ApJS..177..189G} for details of the algorithms used) at the 99\% significance during an earlier observation of the SMC on 1999 May 11 (MJD 51309.668). Details of this observation can be found in Table \ref{tab:ObsLog} . SXP 2.16 was not identified during this observation as a blind period search is heavily affected by the presence of SMC X$-$1 in the \rxte\, field of view. Only with prior knowledge of the detected periodicity can the signal be found in the archival data. We show the pulse profile obtained from this observation in the top panel of Fig. \ref{fig:XTE_phasefold}. We note that the count rate in this profile is much greater than that observed in the lower panel, taken from the observation on 2003 January 05. This is almost certainly due to the presence of SMC X$-$1 in the field of view and the fact \rxte\, was not an imaging telescope. The pulse profile, however, should be unaffected by the presence of the 0.7\,s pulse period of SMC X$-$1. Comparing the two profiles, we can see the depth of modulation of the  1999 observation is $\sim$0.8 counts\,s$^{-1}$, compared with $\sim$0.3 counts\,s$^{-1}$ for the discovery observation of 2003. Unfortunately, due to the presence of SMC X$-$1, we cannot quantify any potential variation in pulse fraction between the two observations. 

\subsection{OGLE results}
\label{sec:OGLE_results}
\begin{figure*}
\begin{center}
\includegraphics[width=\textwidth,natwidth=1091,natheight=839]{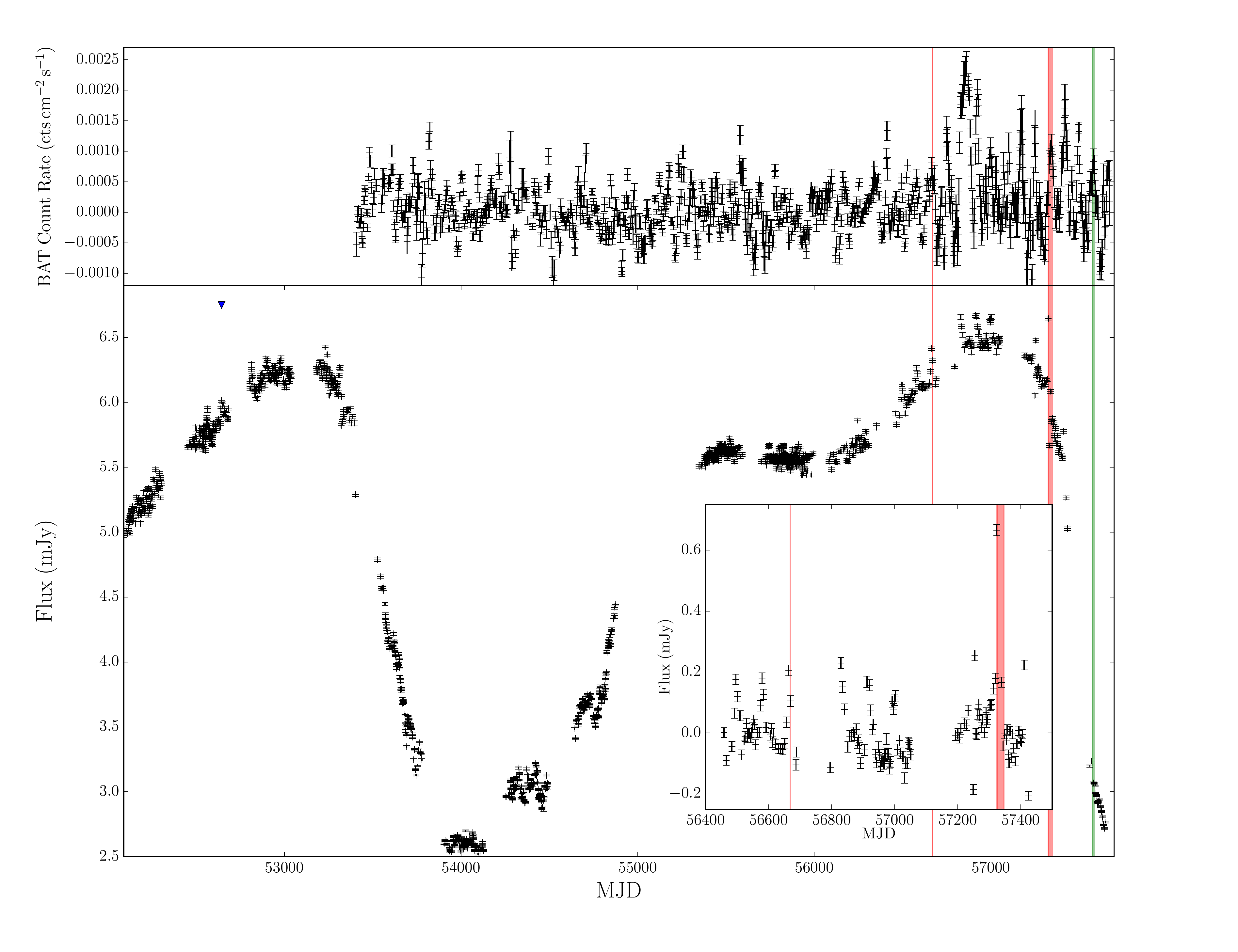}
\caption{Top Panel: Swift BAT 15--50\,keV light curve of \ThisSource. The light curve is a combination of the transient monitor data and the back-processed data as outlined in Section \ref{sec:swift_data}. The red vertical shaded regions represent the dates of the 2014 and 2015 outbursts as detected by \integral\, and the green shaded region is the outburst detected between MJD 57575 -- 57584 as part of the S-CUBED \swift/XRT monitoring. Bottom Panel: OGLE--III and IV combined {\it I}-band light curve of the optical counterpart of \ThisSource\, from MJD 52090.430 -- 57560.425. The blue triangle represents the RXTE observation on MJD 52644.399 during which the spin period of 2.165\,s was first detected. Inset Panel: Zoom of OGLE--III and IV light curve between MJD 56400 -- 57500, detrended for seasonal variations. The red vertical regions correspond to the detected \integral\, outbursts. The $\sim$84\,day variation is clearly seen.}
\label{fig:OGLElc}
\end{center}
\end{figure*}


Fig. \ref{fig:OGLElc} shows the OGLE Phase III and IV combined {\it I}-band light curve of the optical counterpart of \ThisSource\, spanning almost 15 years from from MJD 52090.430 to 57647.221. In this plot, the light curve time stamps and magnitudes have been converted to MJD and milliJansky (mJy) respectively. The RXTE detection of the 2.165\,s spin period is shown as a blue triangle, the \integral\, detected outbursts in 2014 and 2015 are marked with red intervals and the 2016 outburst observed as part of the S-CUBED survey is shown as a green interval. The counterpart shows large variations over the course of the 15 year observations of approximately 4\,mJy, corresponding to $\sim$1 magnitude. On top of this behaviour, recurrent outbursts occurring every $\sim$84\,d can be seen during brighter epochs, for example around MJD 53000 and MJD 56500 onwards. 

\begin{figure}
\begin{center}
\includegraphics[width=0.5\textwidth,natwidth=576,natheight=432]{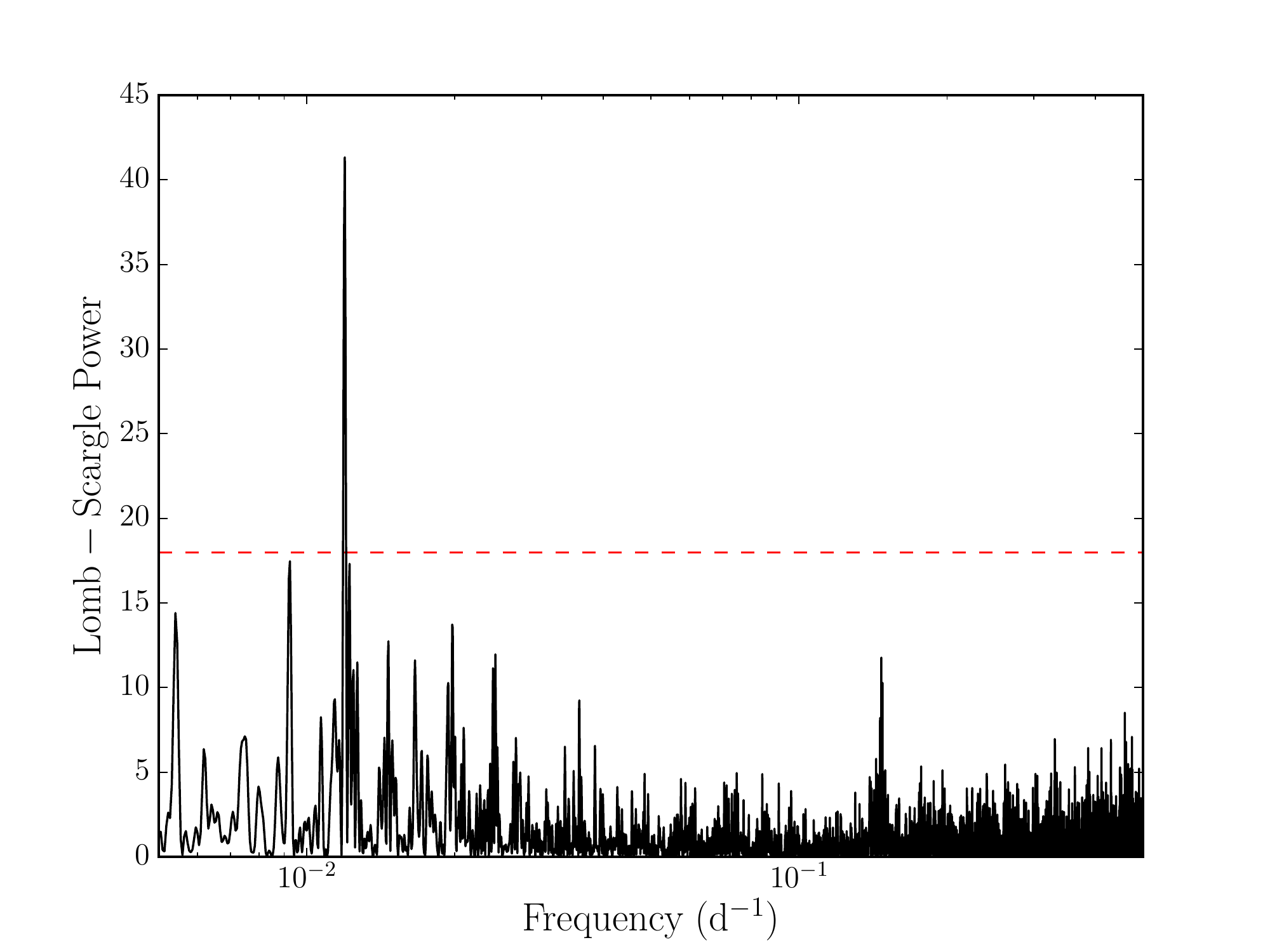}
\caption{Lomb-Scargle Periodogram of combined OGLE III and OGLE IV detrended light curve searched in the range 2--200\,d. The peak in the Lomb-Scargle power is located at 83.67\,d. The red dashed line indicates the 99.99\% confidence level of detection. }
\label{fig:OGLE_LS}
\end{center}
\end{figure}

The combined OGLE III and OGLE IV detrended light curve was tested for the presence of periodicities in the range 2--200\,d by calculating the Lomb-Scargle periodogram using the fast implementation of \citet{1989ApJ...338..277P}. Significance levels, above which a detection would be considered to be true, were calculated using Monte-Carlo simulations and adopting the method of \citet{2005A&A...439..255H} with 100,000 iterations to asses the confidence levels of a signal detection. The resulting periodogram is shown in Fig. \ref{fig:OGLE_LS} with a peak in the Lomb-Scargle power at P=83.67\,d with a significance greater than 99.99\% shown by the red dashed line. In order to ascertain the uncertainty of this signal, we created 10,000 simulated light curves using the bootstrapping method and calculated the periodogram for each. The peak period was recorded for each and the resulting distribution of peak periods was fit using a Gaussian distribution, of which, the standard deviation is taken to be the error on the periodic signal. From this analysis, we derive the period and associated error to be P=$83.67\pm0.05$\,days. 

We also searched each season of data across OGLE III and IV observing campaigns for the presence of  periodicities such as Non-Radial Pulsations (NRPs) as have been seen previously in optical companions in BeXRBs. We carried out the same Lomb-Scargle analysis detailed above, though changed the range over which periodicities were tested from 2--200\,d to 1--200\,d. As the OGLE data are not evenly sampled, defining a Nyquist frequency is difficult, though the approximate sampling is $\sim$1 day leading potentially to strong aliasing effects in the periodogram. Significant detections of signals are found in Seasons 1, 10 and 11 of the combined OGLE III and IV observations. 

The Lomb-Scargle periodogram of OGLE observations was taken between MJD 52090.429 and 52303.095 as part of the OGLE III campaign. reveals a peak at 1.167\,days with a detection significance greater than 99.9\% and also a strong aliasing feature at 6.851\,d. Due to the strong aliasing effects, we cannot be certain which is the true sub-orbital signal. We focus on the signal at 6.851\,d to derive an error and evaluate any evolution in these signals. A bootstrapping analysis, as detailed above, was carried out to evaluate the uncertainty of this signal at 6.851\,d, resulting in a period of P$\,=\,6.851\pm0.024$\,d

Running the same analysis on data taken as part of the OGLE IV campaign between MJD 55699.404 and 55993.014 results in periodic signals at 1.171 and 6.801 days with significances greater than 99.99\%. A Bootstrapping analysis of this observation period leads to a distribution of peak peaks with mean, $\mu=6.880$ and standard deviation $\sigma=0.024$. We therefore report the periodic signal as 6.880$\pm$0.024 days for this observing season. 

\begin{figure}
\begin{center}
\includegraphics[width=0.47\textwidth,natwidth=576,natheight=432]{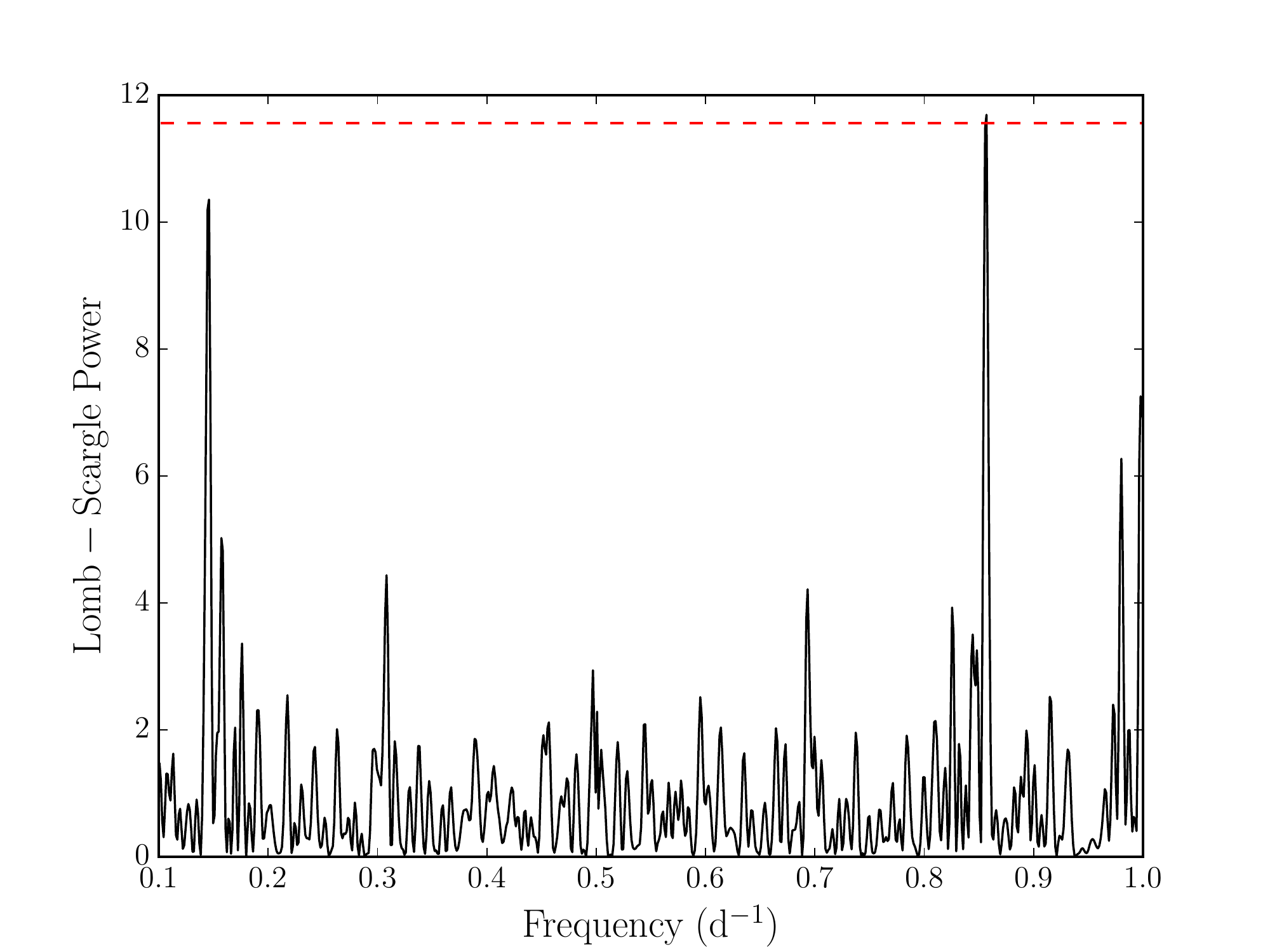}
\includegraphics[width=0.47\textwidth,natwidth=576,natheight=432]{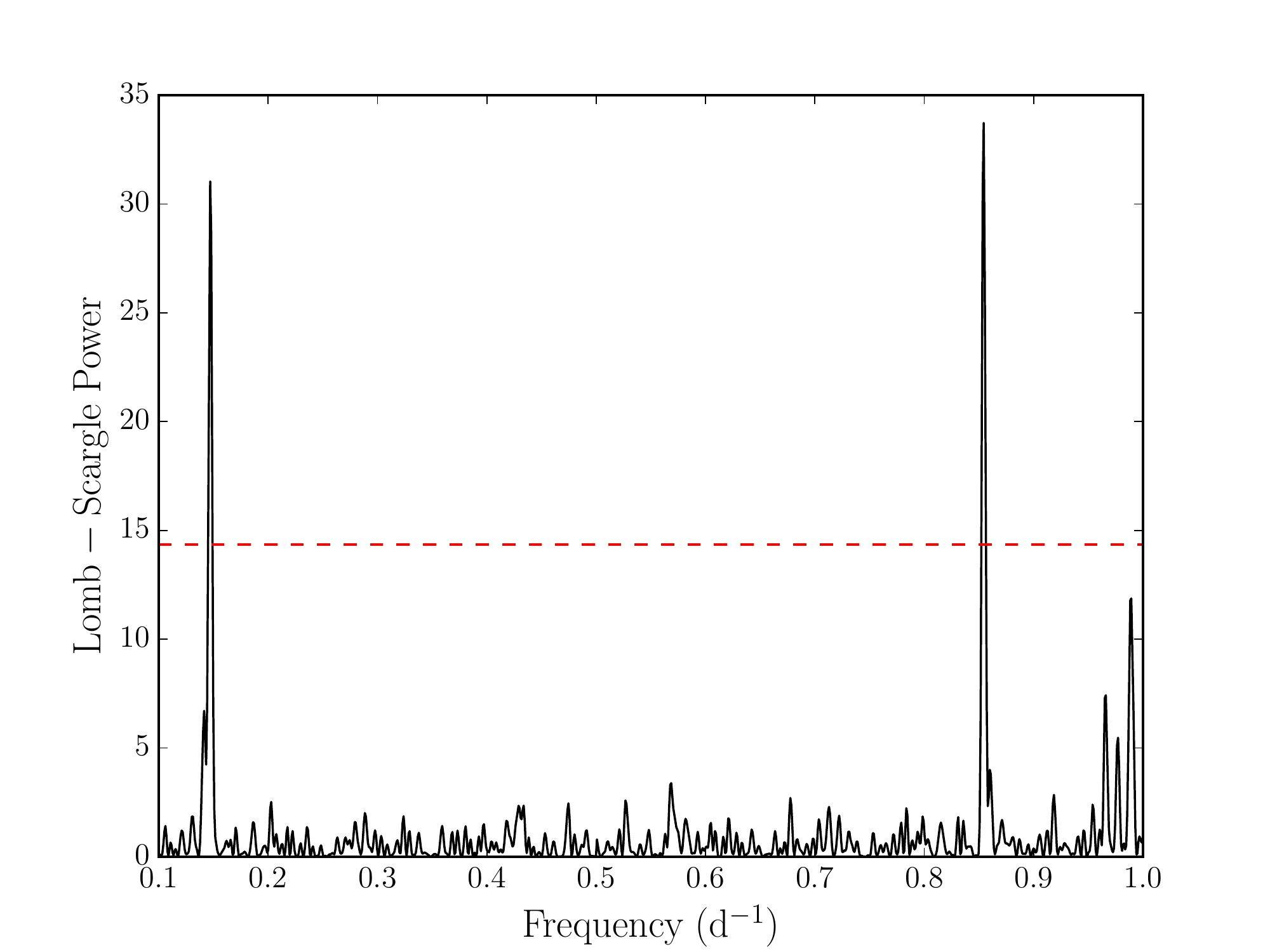}
\includegraphics[width=0.47\textwidth,natwidth=576,natheight=432]{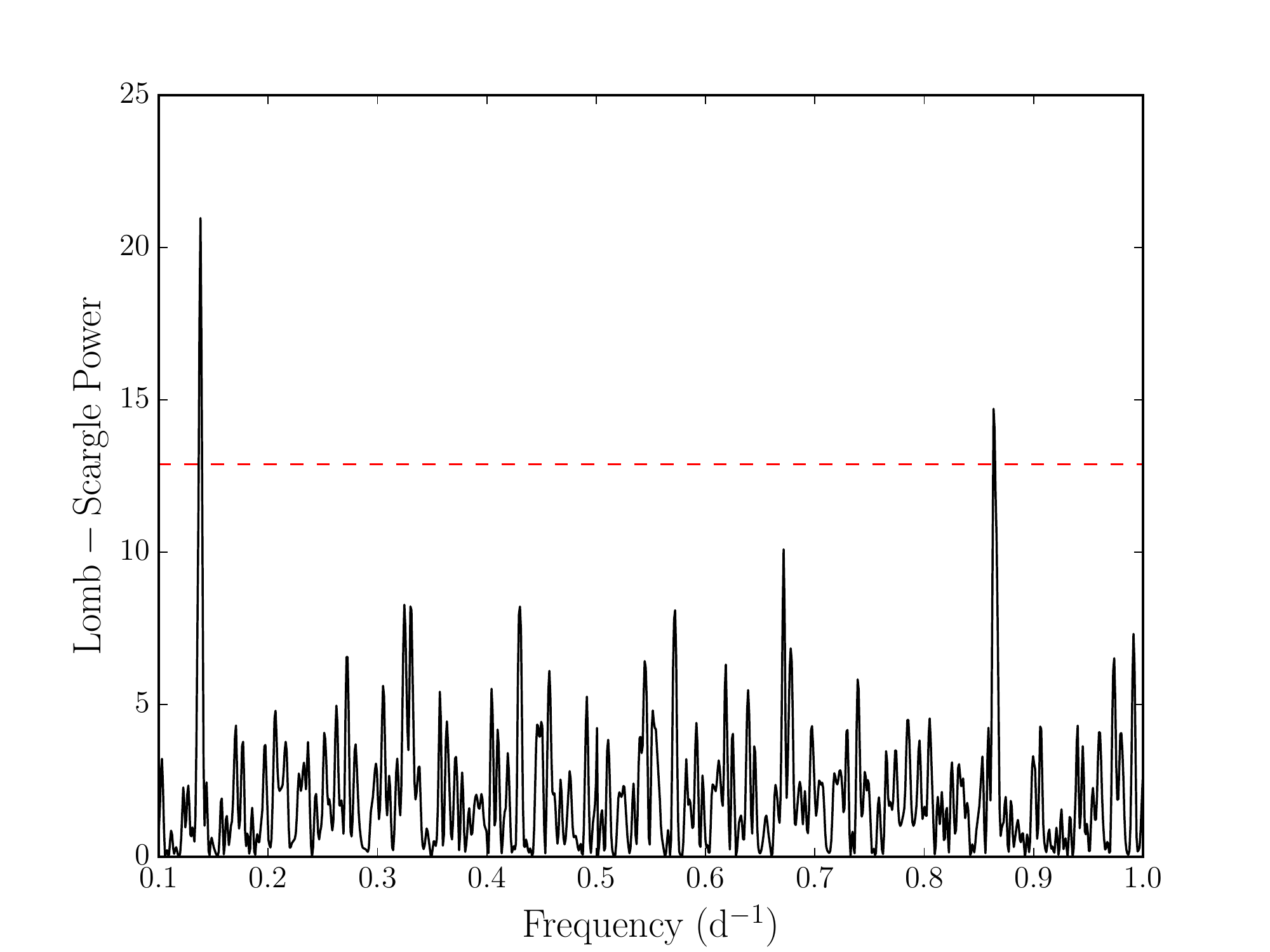}
\caption{Top panel: Lomb-Scargle periodogram of OGLE data taken between MJD 52090.429 -- 52303.095. The resulting periodogram shows signal at 1.167\,days with a detection significance greater than 99.9\% (red dashed line) and also a strong aliasing feature at 6.851\,d. Middle panel: As top panel for observations taken between MJD 55699.404 -- 55993.014. Periodic signals at 1.171 and 6.801 days are detected with significances greater than 99.99\%. Bottom panel: As top panel for observations taken between MJD 56083.43167 -- 56353.022. Periodic signals at 1.158 and 7.237 days are detected with significances greater than 99.99\%}
\label{fig:OGLE_season_ls}
\end{center}
\end{figure}

The resulting periodogram of observations taken between MJD 56083.43167 -- 56353.022 reveals periodic signals at 1.158 and 7.237 days detected with a significance greater than 99.99\%. Bootstrapping analysis of the 7.237 day signal results in a distribution with standard deviation $\sigma=0.008$ days. We therefore report a signal of 7.237$\pm$0.008\,days from this observation period. The periodograms of the three seasons with significant detections and appropriate confidence levels are shown in Fig. \ref{fig:OGLE_season_ls}. 

In order to investigation a possible evolution of the short term periodicities over the course of the observations between MJD 55699.404 -- 55993.014 towards the detected period observations between MJD 56083.43167 -- 56353.022, we subdivided the former observation period into 4 sections with equal numbers of observations in each. We performed the same Lomb-Scargle analysis of these sections as described above, though in this case, due to the much shorter total observation time in each section, we restricted the period search to 1--25\,days. The results of this analysis can be found it Table \ref{tab:OGLE_subsec}. In the first three subsection, two periodicities are significantly detected, the first at around 1 day and the other between 6 and 7 days. In the fourth subsection, a periodicity of 6.873\,days is detected, though no significant signals around 1 day are found. For consistency, we therefore limit our uncertainty calculations to the $\sim$6\,day periodicities, which are also given in Table \ref{tab:OGLE_subsec}.

\begin{table}
\begin{center}
\begin{tabular}{cccc}
\hline
\hline
Section & MJD 			 & Periodicity (d) 			& Significance  \\
\hline
1 		& 55699 -- 55803 & 1.17, 6.77$\pm$0.04		& 99\%			\\
2		& 55803 -- 55853 & 1.02, 6.57$\pm$0.05		& 99.9, 90\%	\\
3		& 55854 -- 55905 & 1.17, 6.78$\pm$0.04		& 99.9\%		\\
4 		& 55906 -- 55993 & 6.91$\pm$0.05			& 99.9\%		\\
\hline
\hline
\end{tabular}
\caption{Results of the Lomb-Scargle analysis of subsections of observations taken between MJD 55699.404 -- 55993.014. In the case of the first three sub sections, two periodicities are significantly detected, one around 1\,day, the other around 6\,days. The fourth section does not show a significant periodicity around 1\,day and hence we restrict uncertainty calculations to the 6\,day periodicities for all subsections. Where two significance values are given, their order denotes their associated periodicity.}
\label{tab:OGLE_subsec}
\end{center}
\end{table}

\section{Discussion}
\label{sec:discussion}
\subsection{Outburst History}
\label{sec:int_discussion}
The first reported outburst of \ThisSource\, detected by \rxte\, coincides with a period of high optical flux and the non-detections three weeks prior and two weeks after suggest that this is a Type I outburst. Unfortunately, there is limited diagnostic information about the pre-discovery outburst from 1999 and so we cannot draw firm conclusions about the nature of this period of activity. 

\ThisSource\, was identified as a transient object during \integral\, observations of the SMC between MJD 56668.038 and 56669.192. During this period the source reached a luminosity of (1.43$\pm$0.23)$\times10^{37}$\lumcgs\, in the 20--40\,keV band and subsequent \integral\, observations revealed no detectable emission from the source. 

The following year, the source was again detected by \integral\, during four observations between MJD 57324.774 and 57347.281. Over the four observations, there is clear evolution in the average luminosity of the source. Initially, between MJD 57324.774 and 57325.874, the source was detected with a luminosity of (4.10$\pm$0.27)$\times10^{37}$\lumcgs\, and shows a general trend of decreasing luminosity. Using the orbital period and ephemeris of Section \ref{sec:BAT_monitorresults}, both outbursts reported here occurred during periastron as determined from the X-ray data and had luminosities of $\sim10^{37}$\lumcgs\,, consistent with Type I X-ray outbursts in Be/X-ray binaries \citep{2011Ap&SS.332....1R}. 

\begin{figure}
\begin{center}
\includegraphics[width=0.5\textwidth,natwidth=725,natheight=535]{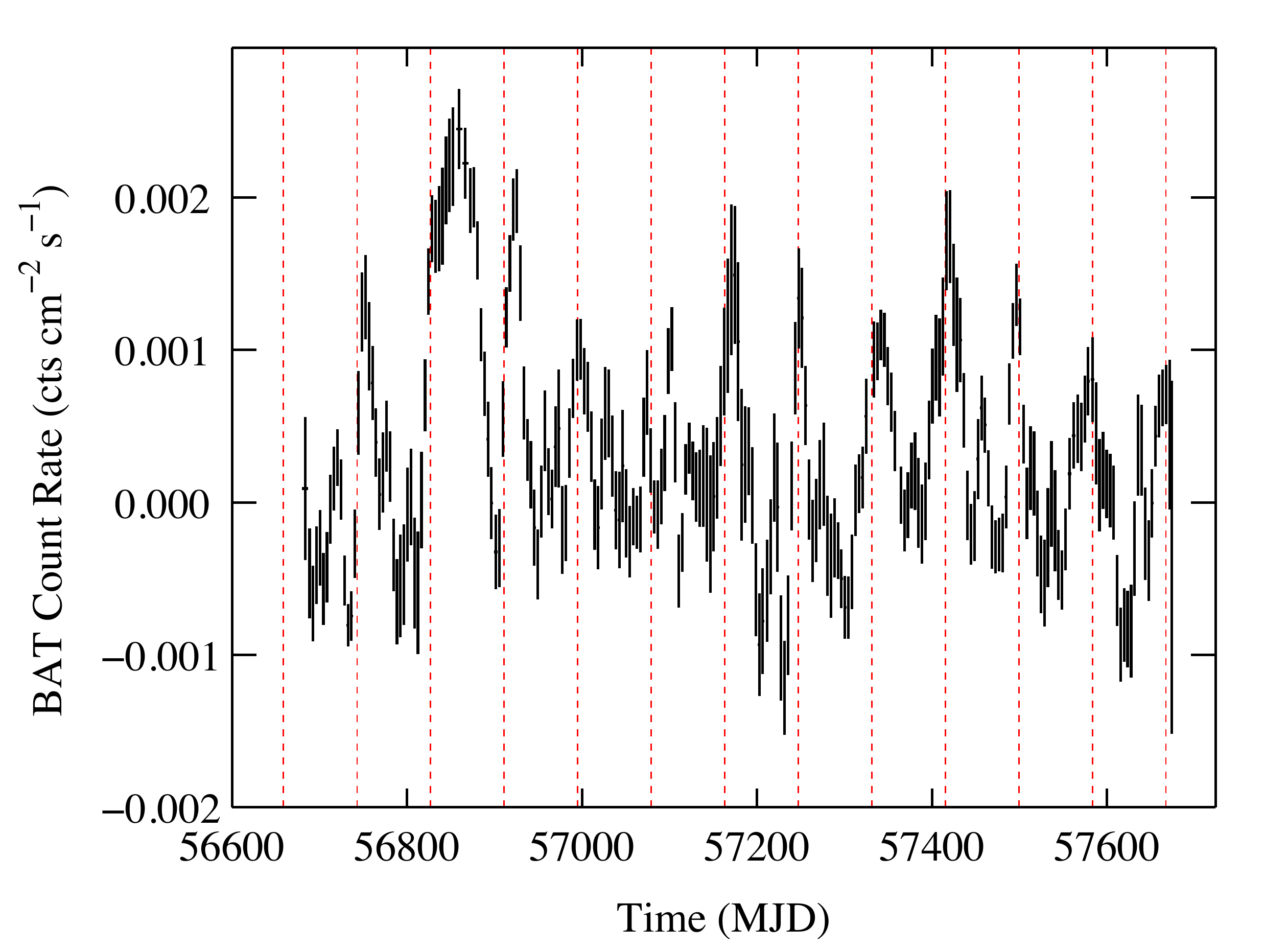}
\caption{\swift/BAT light curve of \ThisSource. The lightcurve is a rebinned and smoothed version of the BAT Transient Monitor one-day resolution light curve. The dashed red lines indicate the times of predicted flux maximum from the 82.5 day period.}
\label{fig:bat_lc}
\end{center}
\end{figure}

The \swift/BAT transient monitor light curve of \ThisSource\, is shown in Fig. \ref{fig:bat_lc}. We also mark in Fig. \ref{fig:bat_lc} the times of the predicted maxima. From the light curve apparent Type I outbursts can clearly be seen during most periastron passages, as well as a potential unreported Type II outburst occurring around MJD 56850. We note that although outbursts can be directly observed at the predicted outburst times, there is cycle to cycle variability in the peak flux and morphology.

The observed outburst detected by \swift/XRT during the S-CUBED survey has phase, duration and luminosity also consistent with being a Type I outburst. 

\subsection{Temporal Analysis}
\subsubsection{Spin Period}
\label{sec:spin_disc}

In Section \ref{sec:RXTE_results}, we present the detection of SXP 2.16 during an observation of the SMC on 2003 January 05. We detect a periodic signal of P$=2.1652\pm0.0001$\,s, which we interpret as modulation of X-rays by a rotating neutron star. We also present the results of archival observations of the SMC on 1999 May 11 where, thanks to knowledge of the spin period as detected in 2003, we now detect a periodicity of 2.16497$\pm$0.00014\,s at the 99\% significance level. Comparing this spin period with the sample of SMC pulsars used by \citet{2014MNRAS.437.3863K} in their study of spin period change in Be/XRBs, we find that SXP 2.16 has one of the fastest spin periods in the SMC (we note that SMC X$-$1 has a spin period of 0.72 seconds, however this is the only known example of a Roche lobe filling supergiant system in the SMC). Comparing this initial detection with the recent measurement of the spin period at $2.16501\pm0.00001$\,s by \citet{2015ATel.8305....1H} suggests a  very small possible spin-up of the pulsar, $\dot{P}\sim1.7\times10^{-5}$ s\,yr$^{-1}$. However, we note that the recent spin period detection of \citet{2015ATel.8305....1H} is only 2$\sigma$ different from the discovery detection period and hence we suggest that this measurement of spin-up is tentative. Such a small change in spin period over the course of $\sim$12.8 years could suggest the neutron star is in spin equilibrium. However, the lack of reported outbursts between the \rxte\, detections and the association with \ThisSource\, suggests the neutron star could have experienced a protracted spin down phase and the inferred spin-up has only occurred since the outburst detected by \integral\, in 2014. In this case, we can set a lower bound of the value of $\dot{P}\sim1.85\times10^{-4}$ s\,yr$^{-1}$ for the possible spin-up rate.
\begin{figure}
\begin{center}
\includegraphics[width=0.5\textwidth,natwidth=576,natheight=432]{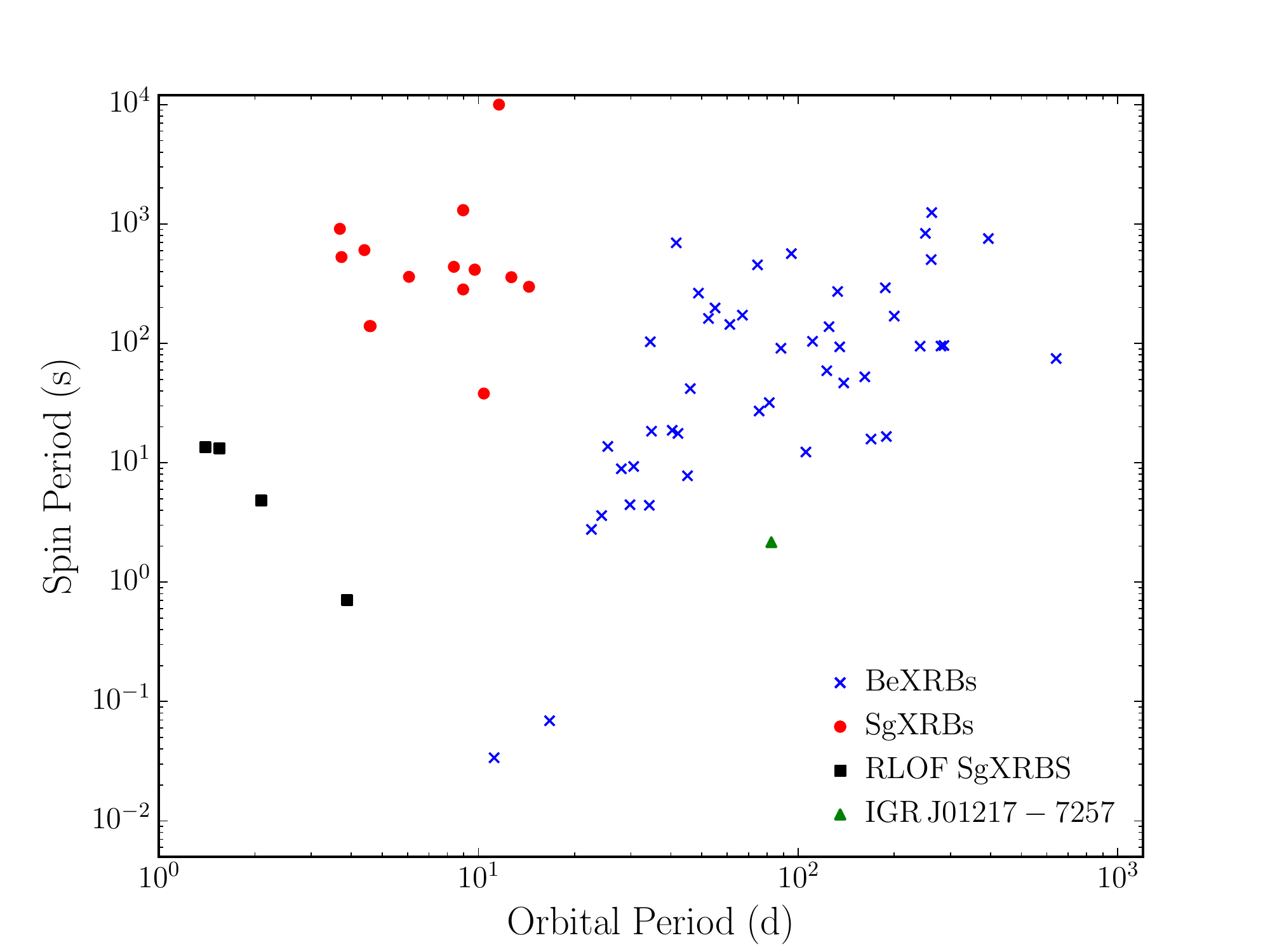}
\caption{P$_{\mathrm{spin}}$--P$_{\mathrm{orb}}$ diagram of HMXBs \citep{1986MNRAS.220.1047C}. Be/X-ray binaries are denoted by blue crosses, wind-fed supergiant X-ray binaries (SgXRBs) by red circles and Roche Lobe Overflow (RLOF) supergiant X-ray binaries by black squares. \ThisSource\, is marked by a green triangle and lies just below the Be/XRB track. Period data taken from \citet{2006A&A...455.1165L} and \citet{2007yCat..34670585B}}
\label{fig:corbet_diag}
\end{center}
\end{figure}

\subsubsection{Orbital Period}
\label{sec:orb_disc}
Using long base-line hard X-ray data taken from the \swift/BAT transient monitor we detect a signal with period, P$=82.5\pm0.7$\,days. This is similar to the periodic signal of P$=83.67\pm0.05$\,days detected in Section \ref{sec:OGLE_results} from a systematic temporal analysis of the light curve taken from both the OGLE III and IV campaigns. While the quoted uncertainties on these detected signals suggest they are different, they are consistent at the 2$\sigma$ level, this is largely driven by the uncertainty in the X-ray period. There are a number of BeXRB systems where there are significant discrepancies between the X-ray and optical periods (see Table 2 and references therein of \citealt{2012MNRAS.423.3663B}). This difference is likely due to the origin of the modulation in each waveband. The X-ray emission and its modulation trace the accretion of material on to the neutron star, which in the case of BeXRBs is when the neutron star is close to the Be decretion disc. The optical modulation, however, is likely due to a combination of the orbital motion of the neutron star around the companion star and the rotation of the decretion disk. The difference in the periodicities here, when compared with other examples of \citet{2012MNRAS.423.3663B}, is relatively small, suggesting that contribution to the signal from the decretion disk motion are small.  In this case we favour the X-ray periodicity as the binary period of this system. 

With the association of SXP 2.16 and \ThisSource\, we can place the source in the P$_{\mathrm{spin}}$--P$_{\mathrm{orb}}$ parameter space. Fig. \ref{fig:corbet_diag} shows the P$_{\mathrm{spin}}$--P$_{\mathrm{orb}}$ diagram of HMXBs \citep{1986MNRAS.220.1047C} with period data taken from \citet{2006A&A...455.1165L} and \citet{2007yCat..34670585B}. We have marked the position of \ThisSource\, assuming the orbital period determined from the X-ray data. The source lies on the bottom side of the Be/X-ray binary track. However, the track is quite broad and hence we do not suggest that this system is peculiar with respect to other systems of the same class. 

\subsubsection{Short period oscillations}

The detection of short period optical oscillations in BeXRBs are often associated with non-radial pulsations of the companion star. In this case, we significantly detect short period oscillations in three epochs of combined OGLE--III and OGLE--IV data. The three epochs all have a similar flux of 5--6\,mJy and occur prior to the onset of flaring behaviour seen in the companion light curve. Non-radial pulsations have been invoked to explain the formation of the decretion disc in BeXRBs (\citealt{2009ApJ...701..396C} and references therein) and hence the onset of Type I outbursts. \citet{2014ATel.5889....1S} also detected short period pulsations (P$=1.173$\,d) in the OGLE-IV light curve searching between outburst times over the whole campaign, in contrast to our approach, where we search individual OGLE observing seasons. Their derived sub-orbital periodicity is consistent with that found during the observations spanning MJD 55699.404 to 55993.014. 

Timing analysis of epochs spanning MJD 55699.404 to 55993.014 and MJD 56083.43167 -- 56353.022 of the OGLE data suggests an evolution of the non-radial pulsation between the seasons. In the case of the $\sim$1\,day signal, we see a shortening of the periodicity, whereas the converse is true for the signals around 6-7\,days.  We suggest that this is tentative evidence for periodicity evolution in non-radial pulsations driving the growth of a decretion disk to a critical size, such that periastron passages of the neutron star generate Type I X-ray outbursts.

Sub-division of the observations between MJD 55699.404 and 55993.014 does not reveal monotonically decreasing/increasing signals and the inferred periodicities are broadly consistent with the periodicity found for the season as a whole.

\subsection{X-ray -- Optical Behaviour}
\label{sec:Xray_Optical}

Fig. \ref{fig:OGLElc} shows the \swift/BAT light curve for both the transient monitor and the back-processed data in the top panel The bottom panel shows the combined OGLE--III and IV light curves spanning $\sim$15 years. In both panels, the \integral\, and \swift\, detected outbursts marked with red and green regions respectively. There is clear association between the detected X-ray and optical outbursts. This is, again, consistent with Type I outbursts from BeXRBs where X-ray outbursts are caused by the interaction of the neutron star with the decretion disk. From the OGLE light curve, we see orbital modulation from the neutron star-decretion disk interactions between the detected \integral\, outbursts. This suggests that due to the observing strategy employed in monitoring the SMC with \integral\,, a number of X-ray outbursts were not observed during the peak in {\it I}-band flux. The observing strategy employed during the monitoring of the SMC is not optimised for the detection of given sources and is instead designed to have observations at regular intervals, restricted slightly be satellite scheduling.

The 11 year period between the \rxte\, detection of the source as SXP 2.16 in 2003 and the re-discovery as \ThisSource\, coincides with protracted low-flux period for the companion star. The lack of X-ray outbursts reported during this time or seen in the \swift/BAT light curve, coupled with the decrease in optical flux suggests that during that phase we are seeing either a reduced decretion disk size, a lack of a decretion disk around the star or observations of the SMC during this period were unfortunately timed. In the first case, the disk has retreated below the radius at which the binary orbital of the neutron star brings it into contact with the decretion outflow. The flux minimum seen in the OGLE light curve around MJD 54000 is then the flux of the companion star combined with that of the decretion disk at its minimum radial extent from the stellar photosphere. In the second case, the flux minimum seen would correspond with the flux from the companion star with no decretion disk. The third case, we believe is unlikely. This is due to the fact, that a similar observing strategy to that used during the flux minimum resulted in the observation of two outbursts in as many years. 

The \swift\, outburst detected during the S-CUBED survey favours the idea that a circumstellar disk is still present at a sufficient size to interact with the neutron star. This X-ray outburst is observed during a period of decreasing {\it I}-band flux, suggesting that during this phase the decretion disk has not sufficiently decreased in size so as to inhibit Type I outbursts. Comparing the behaviour of the companion with the activity seen around MJD 53700, we appear to be seeing the onset a second flux minimum of the companion star with an apparent recurrence time of 10.2 years. Further monitoring of \ThisSource\, during the S-CUBED survey will reveal whether or not X-ray outbursts continue during the expected flux minimum. Detection of Type I outbursts during this time will confirm the presence of a decretion disk.

The analysis shown in Fig \ref{fig:bat_rvt} and \ref{fig:bat_pvt} further suggest that the Type I outbursts cease or are of low enough luminosity not to be detected by \swift/BAT during the period of low optical flux. Including this portion of the \swift/BAT light curve does not enhance the detection of the orbital modulation in the hard X-ray. 
\section{Conclusions}
\label{sec:Conclusions}
We present multi-wavelength observations of the Small Magellanic Cloud Be/XRB, \ThisSource\,, also known as SXP 2.16, during three outburst epochs as well as the original \rxte\, discovery data for the first time. Initially detected on 2003 January 05 during \rxte\, scans of the SMC with a periodicity of 2.1562\,s, the source was not subsequently detected for 11 years (corresponding with a minimum in flux activity from the companion star) until January 2014 when it was detected by \integral\,. Further X-ray outbursts were detected by \integral\, between 2015 October 29 and 2015 November 21 and by \swift\, as part of the S-CUBED survey between 2016 July 06 and 2016 July 15 . We perform temporal analyses of long base line \swift/BAT and OGLE light curves revealing a small discrepancy between inferred orbital periods of 82.5$\pm$0.7 and 83.67$\pm$0.05\,days, respectively. Interpreting the X-ray periodicity as indicative of binary motion of the neutron star, we find that outbursts detected by \integral\, and \swift\, between 2014 and 2016 are consistent with Type I outbursts seen in Be/XRBs, occurring around periastron. Comparing these outbursts with the OGLE data, we see a clear correlation between outburst occurrence and increasing {\it I}-band flux. An periodic analysis of subdivisions of OGLE data reveals three epochs during which short periodicities of $\sim$1\,day are significantly detected which we suggest are non-radial pulsations (NRPs) of the companion star. These seasons immediately precede those exhibiting clear outburst behaviour, suggesting and association between the NRPs, decretion disk growth and the onset of Type I outbursts as has been suggested before.  

\section*{Acknowledgements}
INTEGRAL is an ESA project with instruments and science data centre funded by ESA member states (especially the PI countries: Denmark, France, Germany, Italy, Switzerland, Spain), Czech Republic and Poland, and with the participation of Russia and the USA. The OGLE project has received funding from the National Science Centre, Poland, grant MAESTRO 2014/14/A/ST9/00121 to AU. This work
made use of data supplied by the UK Swift Science Data Centre at the University of Leicester. CMB is supported by the UK Science and Technology Facilities Council and thanks A. B. Hill for fruitful discussions. PAE acknowledges UKSA support. JAK acknowledges support from the NASA grant NAS5-00136. This work was supported in part by NASA grant 14-ADAP14-0167.



\bibliography{Master_references.bib}

\label{lastpage}

\bsp	
\label{lastpage}
\end{document}